\documentclass{aastex631}
\usepackage{xcolor}

\shorttitle{Age Metallicity Relation}
\shortauthors{Wang C. et al.}
\graphicspath{{./}{figures/}}
\begin{document}

\title{Non-monotonic Age–metallicity Relations Across the Galactic Disk Revealed by APOGEE Red Clump Stars}

\author[0000-0002-8713-2334]{Chun Wang}
\altaffiliation{Corresponding authors (wchun@tjnu.edu.cn; huangyang@ucas.ac.cn)}
\affiliation{Tianjin Astrophysics Center, Tianjin Normal University, Tianjin 300387, People's Republic of China.}

\author[0000-0003-3250-2876]{Yang Huang}
\altaffiliation{Corresponding authors (wchun@tjnu.edu.cn; huangyang@ucas.ac.cn)}
\affiliation{School of Astronomy and Space Science, University of Chinese Academy of Sciences, Beijing 100049,  People's Republic of China.} 
\affiliation{National Astronomical Observatories, Chinese Academy of Sciences, Beijing 100012, P.R. People’s Republic of China.}

\author[0000-0003-2471-2363]{Haibo Yuan}
\affiliation{School of Physics and Astronomy, Beijing Normal University, Beijing 100875, People’s Republic of China.}

%\author[0000-0003-3864-068X]{Lizhi Xie}
%\affiliation{Tianjin Astrophysics Center, Tianjin Normal University, Tianjin 300387, People's Republic of China.}

\begin{abstract}

Using APOGEE red clump stars and open clusters with precise ages, we investigate the age--metallicity relation (AMR) across the Galactic disk as a function of Galactocentric radius $R$ and vertical height $|Z|$. For stars older than $\sim$6--8 Gyr, the AMR shows little dependence on either $R$ or $|Z|$ within the uncertainties, suggesting rapid and well-mixed chemical evolution in the early disk. In contrast, younger populations exhibit a clear negative radial metallicity gradient, with the inner disk being more metal-rich than the outer disk, consistent with inside-out disk growth since $\sim$6--8 Gyr ago.
Most importantly, at fixed location, the AMR for ages $\lesssim6$--8 Gyr is non-monotonic: [Fe/H] decreases toward younger ages, reaches a minimum at intermediate ages, and then rises again. Beyond $R\sim11$\,kpc, the relation becomes even more complex, exhibiting an additional reversal and a distinct metallicity peak at an age of $\sim2$\,Gyr. Such complex behavior cannot be explained by continuous local enrichment alone, but instead points to the combined effects of a multi-phase gas-accretion history and radial migration. Episodes of metal-poor gas inflow can dilute the interstellar medium and reset its chemical evolution, while subsequent enrichment raises the metallicity again; radial migration further reshapes the observed AMR by mixing stellar populations formed at different radii and metallicities.
A toy model incorporating inside-out growth, two gas-inflow episodes, and radial migration qualitatively reproduces the observed trends, highlighting the combined roles of episodic gas accretion and radial migration in shaping the chemical evolution of the Galactic disk.

\end{abstract}

\section{Introduction}
%As most of the MW's baryons and angular momentum are  in the disc,  the knowledge of how the disc was  formed  and assembled is a key to understanding  the formation and evolution of the MW as whole.  However,  many questions regarding the disc   formation  and assemblage  remain open, and there are extensive discussions about this topic \citep{Freeman2002,Rix_review,vander2011}. 

The Milky Way, as a typical disk galaxy, offers an exceptional laboratory for constraining galaxy formation and evolution models. Its key advantage is resolvability: we can obtain detailed stellar parameters, including 3D spatial positions and velocities, ages, and chemical abundances, for individual stars. These combined data serve as a powerful chemo-dynamical tag, linking stars to their birth environments and subsequent orbital evolution. The wealth of data from massive spectroscopic surveys: such as Gaia \citep{gaiaedr3,gaiaxp}, APOGEE \citep{apogee1,apogee2,apogee}, LAMOST \citep{lamost,lss-gac}, GALAH \citep{galah}, and DESI \citep{desi2016,desi2}, together with the framework of Galactic archeology, enables us to dissect the assembly history of the Milky Way with remarkable precision.

For low- and intermediate-mass stars, the photospheric abundances of many elements largely reflect the composition of the natal interstellar medium (ISM), although some surface abundances can be altered by internal mixing processes such as dredge-up.
For these conserved elements, the photospheric abundance primarily reflects the composition of the ISM at the time and location of their birth.  In this sense, stellar ages and chemical abundances offer complementary information: ages trace when stars formed, whereas chemical abundances encode the enrichment state of the ISM from which they formed.

The chemical evolution of the Galactic disk is governed by the star-formation history (SFH), together with gas infall, outflow, and recycling processes \citep{Larson, Colavitti2008, Pezzulli2016, Andrews2017, Toyouchi2018, Grisoni2017, Grisoni2018}.  In the simplest picture, the accretion of pristine (metal-poor) gas coupled with ongoing metal enrichment produces a smooth secular rise in the interstellar medium (ISM) metallicity. This produces a theoretical age--metallicity relation (AMR), in which younger stars are systematically more metal-rich than older stars. The AMR therefore offers a direct constrains on the chemical evolution and gas-accretion history of the Galactic disk.
%The star-formation history (SFH) of the Galactic disc, together with gas infall and outflow \citep{Larson, Colavitti2008, Pezzulli2016, Andrews2017, Toyouchi2018, Grisoni2017, Grisoni2018}, imprints a characteristic chemical enrichment pattern on the ISM. 
%This establishes a theoretical age–metallicity relation (AMR), in which younger stars tend to be more metal-rich than older ones. \textbf{In reality, however, this relation is modified by dynamical evolution, most notably radial migration driven by non-axisymmetric structure, such as spiral arms and the Galactic bar. This process can relocate stars far from their birth radii and mix populations that formed under different enrichment conditions \citep{sellwood, Roskar2008, Schonrich, Loebman2011, Haywood2016}.} Consequently, the observed AMR reflects not only the time-dependence of chemical enrichment but also the efficiency and imprint of radial migration, making it a potential diagnostic of disc dynamical evolution.

However, this simple picture is challenged by observations.  The first empirical evidence for an AMR in the Galactic disk was provided by \cite{Twarog1980}, based on Strömgren photometry of F-type stars, and was consistent with the classical expectation of a smoothly enriching ISM. However, \cite{Edvardsson1993} found  no strong AMR in the solar neighborhood. They found the local AMR exhibits a large metallicity dispersion at nearly all ages, challenging the assumption that stars of a given age should have nearly identical metallicities.  These trends  have since been found in several other studies of nearby stars \citep[e.g.,][]{Bensby2004, Haywood2013, Feuillet2016, Xiang2017, Wu2019}. \cite{Edvardsson1993}  interpreted the metallicity scatter at a given age as
an indication that these stars were born in different regions of the
Galaxy with distinct chemical enrichment histories \citep[e.g.,][]{Bensby2004, Casagrande2011}. With the advent of large-scale stellar surveys: such as Gaia, LAMOST, APOGEE, GALAH, and DESI,  investigations of the AMR have been extended from solar neighborhood to larger Galactic disk region. \cite{Feuillet2019} found the AMR  varies with both Galactocentric radius and distance from the disc mid-plane, suggesting that radial migration plays a significant role in shaping the present day disc. Using turnoff stars selected from GALAH\,DR3 \citep{Buder2021}
and Gaia\,EDR3 \citep{Gaia2021}, \cite{Sahlholdt2022} further inferred a three phase formation history of the Milky Way through studying the age metallicity distributions.
%{Haywood2013, Feuillet2016, Xiang2017, Wu2019} showed instead that the local AMR exhibits a large metallicity dispersion at nearly all ages, challenging the notion of a tight one-to-one AMR.  This observation implies that stars currently located in the same region did not necessarily form from the same positions or follow a single local enrichment history.

Meanwhile, theoretical chemical evolution models have suggested that gas accretion can generate distinct enrichment pathways across the Galactic disk \citep{Chiappini1997}. In particular, the accretion of relatively metal poor gas may dilute the pre-existing interstellar medium, thereby modifying both the gas phase metallicity and the subsequent chemical enrichment history.  The inside-out growth of the Galactic disk is associated with different star formation histories and enrichment rates in the inner and outer disk. These differences are reflected in negative radial metallicity gradients \citep{Minchev2018,Wang2019}, systematic radial variations in the stellar metallicity distributions \citep{Frankel2019}, and a gradual increase in disk scale lengths over time \citep{Mackereth2017}. Nevertheless, chemical evolution alone is insufficient to explain the substantial scatter observed AMR. Radial migration, driven by non-axisymmetric structures such as spiral arms and the Galactic bar, must therefore also be taken into account.  Radial migration operates through both ``churning'', which changes a star’s guiding-center radius, and ``blurring'', which increases the radial extent of its orbit \citep{sellwood, Roskar2008, Schonrich, Loebman2011, Haywood2016}. As a result, stars with similar ages but markedly different metallicities can coexist in the same spatial region at the present day. This creates a fundamental ambiguity in interpreting the AMR.  This introduces a fundamental ambiguity in interpreting the AMR. Although the AMR encodes the temporal evolution of the local interstellar medium, the observed relation is also reshaped by stars that migrated from regions with different chemical enrichment histories. The AMR therefore cannot be regarded as a purely local record of chemical evolution without explicitly accounting for dynamical mixing.

The key difficulty is that these effects are strongly degenerate when the AMR is measured only in the solar neighbourhood. A large age metallicity scatter may indicate inefficiently mixed or spatially varying chemical evolution, radial migration from different birth radii, or a combination of both. Consequently, determining how the AMR varies across the Galactic disk is essential for separating chemical evolution effects from dynamical mixing and for constraining the underlying gas accretion and star formation histories.

%Motivated by this tension and by the need to disentangle chemical evolution from dynamical redistribution, we investigate the spatial variation of the AMR across the Galactic disk using a red clump stellar sample selected from APOGEE DR\,17 \citep{Abdurrouf2022}, for which reliable ages and chemical abundances are available. 
%Abundance ratios offer important complementary diagnostics of stellar formation and enrichment histories. Recent studies have suggested that stellar age correlates more tightly with abundance ratios such as [$\alpha$/Fe] than with overall metallicity \citep{Haywood2013, Bensby2014}. [Al/Fe] may provide further diagnostic power, as aluminium is an odd-$Z$ element whose production is strongly linked to massive stars, with yields that depend sensitively on the metallicity of the progenitor. \textbf{Aluminium is released primarily through core-collapse supernovae, and its abundance also shows a measurable dependence on stellar age, as found in previous studies \citep{Nissen2015, Feuillet2018, Bedell2018}.} We therefore use [Mg/Fe] and [Al/Fe], together with [M/H], to help distinguish stellar populations with different formation and enrichment histories. \textbf{Here, we use [M/H] rather than [Fe/H] because [M/H] reflects the overall metal content, whereas [Fe/H] traces only the iron abundance. Thus, the age–[M/H] relation provides a more representative view of the Galactic disk’s chemical enrichment history.}

Motivated by this tension and by the need to disentangle chemical evolution from dynamical redistribution, we investigate the spatial variation of the AMR across the Galactic disk using a red clump stellar sample selected from APOGEE DR\,17 \citep{Abdurrouf2022}, for which reliable ages and chemical abundances are available. We focus on the age--[Fe/H] relation because it provides a direct and physically meaningful probe of the chemical enrichment history of the Galactic disk.  In this work, we therefore use the age--[Fe/H] relation to investigate the AMR spatial variations and to constrain the relative contributions of chemical enrichment and stellar dynamical redistribution to the present day structure of the Galactic disk.

%Despite the importance of the AMR, its spatial variation across the Galactic disk has not yet been systematically characterized in sufficient detail. Such a study can provide stronger constraints on the star formation history, gas accretion and outflow, and radial mixing processes that have shaped the Milky Way. In this paper, we investigate the AMR in different regions of the Galactic disk using a red clump stellar sample selected from APOGEE DR\,17\citep{Abdurrouf2022}, for which reliable ages and chemical abundances are available.

%Despite the importance of the AMR, its spatial variation across the Galactic disk has not yet been systematically characterized at sufficient resolution. Mapping how the AMR changes with Galactocentric radius and height can place stronger constraints on the disc’s star-formation and gas-flow history, while directly probing the cumulative imprint of radial mixing. In this paper, we investigate the AMR across different regions of the Galactic disk using a red clump sample selected from APOGEE DR17 \citep{}, for which robust age estimates and high-precision chemical abundances are available.

The paper is organized as follows. In Section 2, we briefly introduce the red clump stellar sample. Section 3 presents the AMR at various spatial locations across the Galactic disk. In Section 4, we discuss the implications of our results for the formation and evolution of the Galactic disk. Finally, Section 5 summarizes our main conclusions.

%chemical tomography of the Galactic disk using a large sample of RGB stars from the LAMOST Galactic surveys  \citep{Wu2019}. The sample stars have robust age estimates, which allow us to explore the age-[Fe/H]-[$\alpha$/Fe] relation as well as the variation of disc spatial structure parameters with chemical abundance and age using a significantly larger sample than previous work. Furthermore, the large sample and the simple target selection function of the LAMOST surveys  allow us to robustly correct for the select effects  \citep[e.g.,][]{Wang2019,Chen2018,Xiangmass}. %The stellar number density in the [Fe/H]-[$\alpha$/Fe] plane at different spatial positions of the Galactic disc

\section{Red clump stellar sample selected from the APOGEE}

Our analysis employs a Galactocentric cylindrical coordinate system. The coordinates are defined as follows: $R$ is the radial distance from the Galactic center; $\Phi$ is the azimuthal angle, measured in the direction of Galactic rotation, with $\Phi=0^{\circ}$ pointing toward the Galactic anticenter; and  $Z$ is the vertical height, with $Z=0$ at the mid‑plane and positive toward the North Galactic Pole.  
We adopt the conventional value of ($R_\sun, Z_\sun$) = (8.34, 0.025)\,kpc for the Sun’s position \citep{Juric2008, Reid2014}.

\subsection{Sample selection}

We selected red clump stars from the APOGEE DR17 Red Clump Catalog \citep{Abdurrouf2022}, which contains 50,387 objects \citep{Bovy2014}. Stellar parameters were derived with the APOGEE Stellar Parameters and Chemical Abundances Pipeline \citep[ASPCAP;][]{Garca2016, Jonsson2020}, and chemical abundances for more than 20 elements were estimated with typical precision of [ X/H]$\leq$0.1\,dex.    Ages for the red clump stars were estimated using an XGBoost model trained on a high quality calibration sample of 3,060 red giant and red clump stars with asteroseismic ages from the combined APOGEE--Kepler data set \citep{Anders2023}. Twenty percent of the data were reserved as an independent test sample, which was used to evaluate the model performance and estimate the statistical age uncertainties. The predicted ages show good agreement with the asteroseismic ages, with statistical age uncertainties below $25\%$. Tests based on duplicate APOGEE DR17 observations further support the reliability of these uncertainty estimates.
By statistically characterizing the distribution of the individual relative age uncertainties in our red clump sample, we find a mean value of $19\%$ and a standard deviation of $3\%$. Nearly all stars have relative age uncertainties below $25\%$, although younger stars tend to have larger relative uncertainties. We therefore adopt the individual age uncertainties in the subsequent analysis. 
Distances were calculated following \citet{Bovy2014}. Although \citet{Anders2023} provide age estimates for both red clump and red giant branch stars, we restrict our analysis to red clump stars. As standard candles, red clump stars provide distance estimates with typical uncertainties of $\sim5$--$10\%$, which can be more reliable than Gaia parallaxes for stars beyond $\sim5$ kpc \citep{Huang2021}. This is particularly important for our analysis of the AMR in the $R$--$Z$ plane. In addition, ages inferred for red clump and red giant branch stars may suffer from systematic differences \citep{Wang2023}. Combining the two evolutionary stages could therefore introduce spatially dependent biases if their relative fractions vary across the Galaxy. We note that a red-clump-only sample may itself be subject to age-selection biases and may not fully represent the underlying Milky Way stellar populations \citep{Ness2016, Girardi2016}.

%Red clump stars serve as standard candles, yielding distance uncertainties of roughly 5–10\%, which can be more reliable than Gaia parallaxes for stars beyond 5 kpc \citep{Huang2021}.

We constructed our sample by applying the following quality and population criteria to the APOGEE DR17 red clump Catalog to ensure a well-defined stellar population and robust parameter estimates:
\begin{itemize}
 \item Spectral signal-to-noise ratio: $\mathrm{SNR} > 100$, to ensure reliable determinations of stellar ages and metallicities. \item Effective temperature: $4000 < T_{\mathrm{eff}} < 5400\,\mathrm{K}$, approximately covering the effective temperature range of giant stars.
 \item Angular momentum along the $Z$-axis: $L_Z > 500\,\mathrm{kpc}\cdot\mathrm{km\,s^{-1}}$, to select disk stars. 
 \item $\mathrm{ASPCAPFLAG} \neq 7$ and $\mathrm{ASPCAPFLAG} \neq 23$. In APOGEE DR17, $\mathrm{ASPCAPFLAG}=7$ indicates an overall warning for a star, whereas $\mathrm{ASPCAPFLAG}=23$ indicates that the star is overall bad. 
 \item Relative age uncertainty: $<30\%$. 
 \item Uncertainties in [Fe/H]: $<0.15\,\mathrm{dex}$. \end{itemize}
This selection yields a final sample of 26,563 red clump stars.

\begin{figure}
\centering
\includegraphics[width=6.5in]{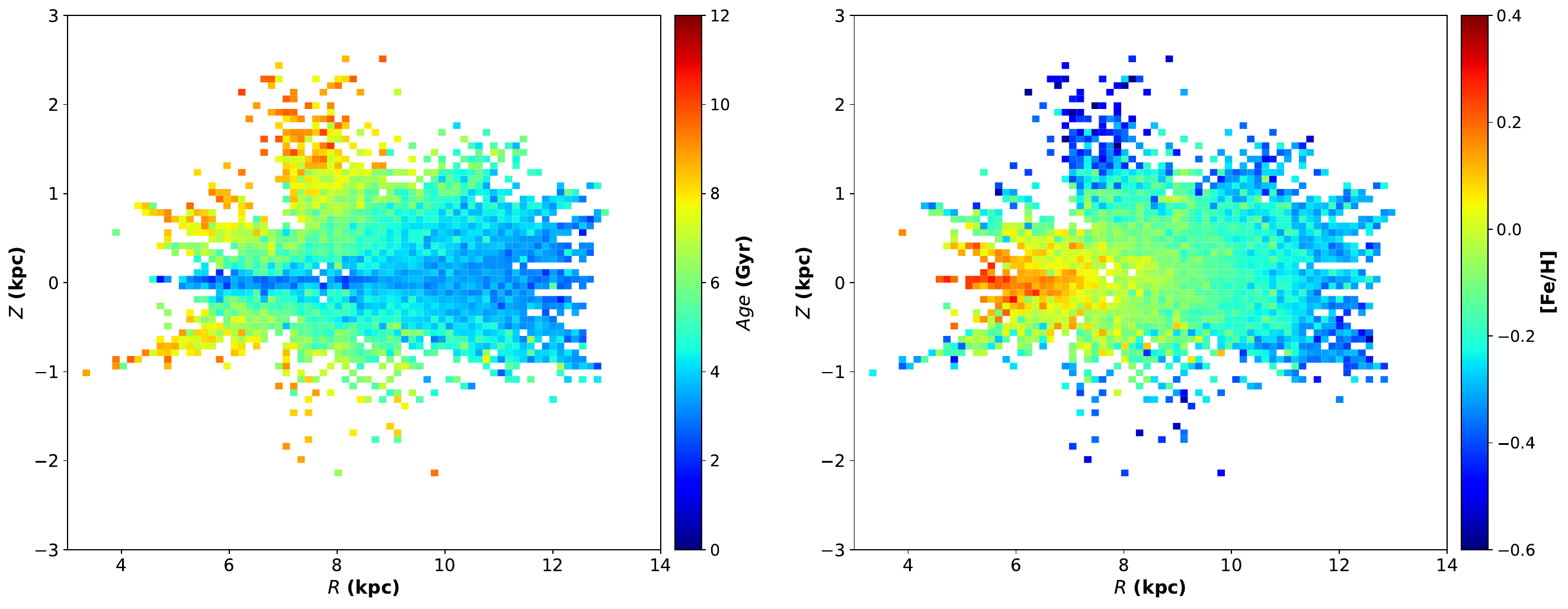}

\caption{Distributions of the final APOGEE red clump sample in the $R$--$Z$ plane, color-coded by the median stellar age (left; $\sim0$--$12~\mathrm{Gyr}$), and the median $[\mathrm{Fe/H}]$ (right; $[\mathrm{Fe/H}]\in[-1.0,+0.5]~\mathrm{dex}$) for all the stars in each bin. The sample spans $R \in [3,14]$~kpc and $|Z| \le 3$~kpc.
}
\label{age_distribution}
\end{figure}

Figure~\ref{age_distribution} summarizes the age, [Fe/H] distributions in the $R$--$Z$ plane of the final red clump sample. It spans Galactocentric radii $R\simeq3$--14\,kpc and vertical heights $|Z|\leq 3$\,kpc. The sample covers ages of $\sim$0--12\,Gyr, metallicities $-1.0\lesssim \mathrm{[Fe/H]}\lesssim +0.5$\,dex, enabling an investigation of the AMR  across a large fraction of the Galactic disk.

\section{Age metallicity relation}
In this section, to quantify the importance of radial migration and gas infall processes in the Galactic disk formation and evolution, we investigate the AMR across the Galactic disk by binning the red clump sample on a two-dimensional ($R$, $|Z|$) grid. We first divide the sample into five Galactocentric radius bins: $3 < R < 5$\,kpc, $5 < R < 7$\,kpc, $7 < R < 9$\,kpc, $9 < R < 11$\,kpc, and $11 < R < 13$\,kpc.  Each radial bin is further subdivided into three vertical-height bins:  
$|Z| < 0.5$\,kpc, $0.5 < |Z| < 1.0$\,kpc,  and $1.0 < |Z| < 2.0$ \,kpc.  Within each spatial bin, we further divide the stellar populations into age bins with a spacing of \(1.0\,\mathrm{Gyr}\), starting at an age of \(0.5\,\mathrm{Gyr}\). We adopt this lower limit because very young stars are relatively rare and their measurements are associated with larger relative uncertainties. For each age bin, we compute the median age and the median [Fe/H]. Bins containing fewer than three stars are excluded to avoid extreme outliers. For each retained age bin, we estimated the uncertainty in the median [Fe/H] using bootstrap resampling of the stars in that bin. Specifically, for a bin containing $N$ stars, we generated $1000 \times N$ bootstrap realizations. In each realization, $N$ stars were randomly drawn with replacement from the original sample in the bin. The measurement uncertainties in age and $\mathrm{[Fe/H]}$ were incorporated by perturbing the corresponding values according to their reported uncertainties. We did not include the uncertainties in $R$ and $|Z|$, because the distance uncertainties of red clump stars are small and the analysis was performed within fixed $R$--$|Z|$ bins. Their effect on the bin memberships is therefore expected to be negligible. The uncertainty in the median $\mathrm{[Fe/H]}$ was then estimated from the distribution of the bootstrap medians.

%\subsection{The relationship between [M/H] and age}

\begin{figure}
\centering

\includegraphics[height=0.8\textheight]{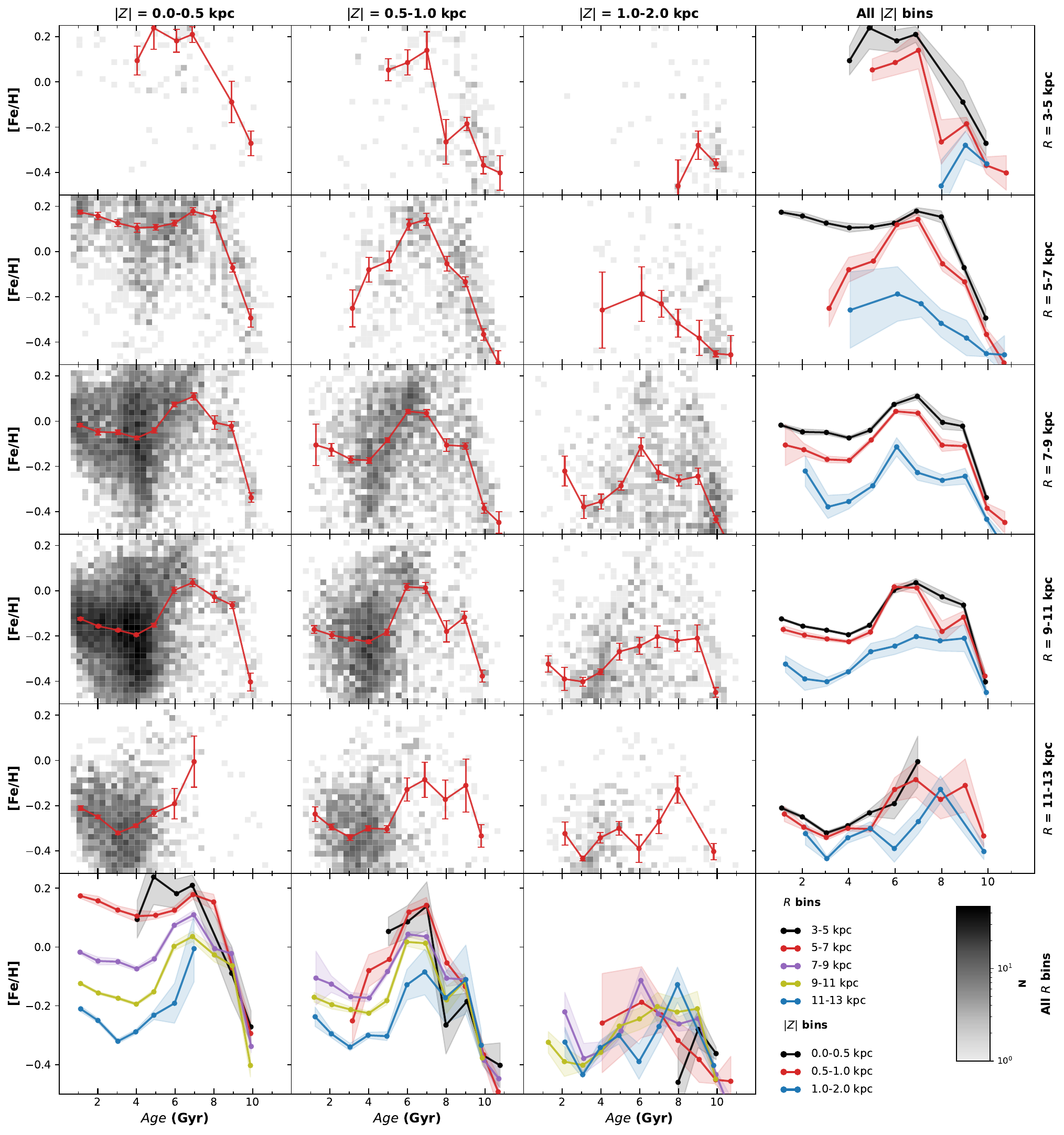}
\caption{The AMR in different Galactic disk bins.
Panels are grouped by five radius bins ($R=3$--5, 5--7, 7--9, 9--11, and 11--13~kpc) and three height bins ($|Z|=0$--0.5, 0.5--1.0, and 1.0--2.0~kpc).
In each $(R,|Z|)$ panel, the grayscale background shows the stellar density distribution, and the red curve with error bars marks the binned median $\mathrm{[Fe/H]}$ versus age.
The ``All $|Z|$ bins'' panels compare the three $|Z|$ bins at fixed $R$, and the ``All $R$ bins'' panels compare the five $R$ bins at fixed $|Z|$; shaded regions indicate the associated uncertainties.}

\label{age_mh}
\end{figure}

Figure \,\ref{age_mh} shows the relationship between age and [Fe/H] across the Galactic disk from $3 < R < 13$\,kpc and $|Z| < 2.0$\,kpc. From this figure, we identify the following features of the AMR:
\begin{itemize}
\item At old ages ($\gtrsim 8$--10~Gyr), $\mathrm{[Fe/H]}$ increases toward younger ages. The trend is broadly similar across all the ($R$, $|Z|$) bins considered.
\item At younger ages ($\lesssim 8$~Gyr), the AMR shows a stronger dependence on position within the disk. At fixed age, stars at smaller $R$ and lower $|Z|$ are systematically more metal rich.
\item In the young regime, the AMR is non-monotonic, with $\mathrm{[Fe/H]}$ decreasing toward younger ages to a minimum before rising again at the youngest ages. The non-monotonic trend is most pronounced at $|Z| < 0.5$\,kpc. Moreover, the stellar age at which the [Fe/H] reaches its minimum depends on spatial location $(R,Z)$, varying strongly with galactocentric radius $R$ but only weakly with height $Z$: it ranges from $\sim 3.0\,\mathrm{Gyr}$ in the outer disk ($R=11$--$13\,\mathrm{kpc}$) to $\sim 5.0\,\mathrm{Gyr}$ in the inner disk ($R=5$--$7\,\mathrm{kpc}$).

\end{itemize}

Previous studies have also investigated the spatial variation of the AMR across the Galactic disc, and have consistently found that the AMR is non-monotonic rather than a simple one-to-one relation \citep{Feuillet2019,Lian2022, Sahlholdt2022}. In particular, the results of \citet{Lian2022} and \citet{Sahlholdt2022} are highly similar to ours: the metallicity initially increases towards younger ages, followed by a rapid decrease at ages of approximately $7$--$9~\mathrm{Gyr}$, reaches a minimum, and then increases again towards younger ages. However, \citet{Lian2022} find that the age corresponding to the metallicity minimum becomes older in the outer disc, which is inconsistent with our result. This discrepancy may be related to the different definitions of radial position adopted in the two studies: we use the Galactocentric radius, whereas \citet{Lian2022} use the guiding centre radius.

\section{Constraining the formation and evolution of the Galactic disk(s)}
%\subsection{ The formation scenario of the Galactic disk at an early epoch}
The AMR of the oldest stellar populations ($\tau > 8$ Gyr) provides important clues to the early formation of the Galactic disk. These old populations exhibit remarkably similar AMRs over a wide range of Galactocentric radii, implying that  the radial metallicity gradients in the early disk were relatively weak or nearly flat,  consistent with findings by \cite{Xiang2015} and \cite{Wang2019}. Previous studies \citep{Brook2004,Brook2005,Haywood2019} have suggested that the Galactic thick disk formed rapidly during the early evolution of the Milky Way. Such a rapid formation phase, possibly driven by a highly turbulent, gas rich merger associated with the progenitor of Gaia-Sausage-Enceladus \citep{Xiang2022}, could have efficiently mixed the interstellar medium (ISM) and erased initial radial metallicity variations. Based on the AMR, \citet{Sahlholdt2022} suggested that the early stellar population formed in a well-mixed, turbulent medium.  \citet{Imig2023} found that the high-$\alpha$ disk has a relatively flat metallicity gradient and a broader metallicity distribution function than the low-$\alpha$ disk, suggesting that radial migration may also have contributed to its present day flat gradient.  A similar picture is suggested by observations of galaxies with $9.0 < \log(M_\star/M_\odot) < 9.5$ at $0.7 \leq z \leq 1.5$.  These galaxies exhibit approximately flat radial profiles of the specific star formation rate (sSFR), which may be related to a self-similar mode of galaxy growth \citep{Nelson2016, Orr2017, Nelson2021}. Taken together with these extragalactic results, our findings favor the interpretation that the star forming gas was already well mixed when the stars formed, although subsequent radial migration may also have contributed to the present day chemical distribution.

As shown in Section\,3, the younger disk population ($\tau<8$\,Gyr) display markedly different AMRs at different Galactocentric radii and heights above the Galactic plane. In particular, mono-age populations  exhibit negative radial metallicity gradients. These trends are consistent with previous studies \citep{Xiang2015,Wang2019}. They can be naturally understood in the context of the “inside-out” disk formation scenario \citep{Matteucci1989,Schonrich2017}, in which the star formation rates (SFRs) in the inner disk exceed those in the outer disk, leading to more rapid chemical enrichment in the inner regions and hence a negative radial [Fe/H] gradient.   

In conclusion, the evolution of the Galactic disk therefore appears to have proceeded through two qualitatively different phases. At early times ($\tau \gtrsim 8\,\mathrm{Gyr}$), the disk evolved in a relatively self-similar and chemically well-mixed manner, producing weak radial metallicity gradients. At later times ($\tau \lesssim 8\,\mathrm{Gyr}$), the growth became increasingly "inside-out", leading to the development of negative radial metallicity gradients.

%In conclusion, the early Galactic disk evolution ($\tau \gtrsim 8\,\mathrm{Gyr}$) favors a self-similar growth mode, whereas the late-time evolution ($\tau \lesssim 8\,\mathrm{Gyr}$) is better described by an inside-out growth scenario.

\subsection{The effects of the radial-migration and gas infall processes}
The classical two-infall model proposes that the Milky Way experienced two major episodes of gas accretion, associated primarily with the formation of the halo--thick disk and the thin disk, respectively \citep{Chiappini1997}. In this framework, the second infall is dominated by relatively metal-poor gas accreted from the intergalactic medium. The arrival of this low-metallicity gas dilutes the pre-existing metal-enriched ISM and temporarily lowers the metallicity of the gas from which new stars form. As star formation continues, star formation and chemical enrichment then raise the metallicity again. The combination of dilution followed by re-enrichment therefore provides a natural explanation for a non-monotonic AMR.

Recent implementations of the two-infall framework offer further support for this interpretation. Based on a Bayesian analysis of the stellar distribution in the [$\alpha$/Fe]--[Fe/H] plane using APOGEE DR16 data, \citet{Spitoni2021, Spitoni2022} inferred a significant delay between the two gas infall episodes associated with the formation of the  thick and thin disks over a broad range of Galactocentric radii. Their models predict a rapid decline in stellar metallicity following the delayed second infall, qualitatively resembling the declining branch of the AMR observed in our data (see, e.g., Figure~14 of \citealt{Spitoni2021}).

However, the metallicity decline in our sample extends over a much longer interval, approximately 3--4\,Gyr, than the $\sim1$\,Gyr decline predicted by the chemical-evolution models of \citet{Spitoni2021}. This difference indicates that gas infall alone is unlikely to account for the full shape of the observed AMR. Instead, dynamical processes must have acted to broaden and reshape the chemical-evolution signature. Radial migration is a particularly natural candidate, because it redistributes stars away from their birth radii and thereby mixes populations formed under different chemical conditions.

The role of radial migration becomes especially clear in the outer disk. In an inside-out formation scenario, the inner disk enriches more rapidly and is more metal-rich than the outer disk at a given time. Radial migration subsequently redistributes stars away from their birth radii. Because older stars have had more time to migrate, metal-rich stars born in the inner disk can contribute increasingly to older populations observed at larger Galactocentric radii. This interpretation is consistent with several observational features, including the presence of super-metal-rich stars whose metallicities exceed those expected from their local environments \citep{Minchev2013, Kordopatis2015, Lehmann2024}, old metal-rich stars in the outer disk \citep{Trevisan2011}, and the skewness of the metallicity distribution functions \citep{Hayden_2015, Wang2019}.   Younger stars, by contrast, have experienced less migration and more closely reflect the lower metallicity of their local birth environment.  
%\textbf{\cite{Lu2022} and \cite{Johnson2021} both pointed out that radial migration can significantly alter the observed AMR, mainly by broadening the metallicity distribution at a given age. The simulations of \cite{Lu2022} suggest that radial migration alone may not necessarily produce the turning point in the AMR; instead, it may be primarily associated with the accretion of low metallicity gas from satellite galaxies and the subsequent formation of new stars. Compared with the inner disc, radial migration has a stronger impact on the outer disc because a larger fraction of outer disc stars have migrated outward from the inner disc.}
Consequently, the observed AMR in the outer disk is not simply a record of the local chemical-enrichment history: it reflects the superposition of stars formed at different radii and subsequently redistributed by Galactic dynamic processes.

We therefore interpret the observed non-monotonic AMR as the combined outcome of two episodic gas accretion \citep{Chiappini1997,Chiappini2001, Spitoni2021}, continuous chemical enrichment, inside-out disk growth \citep{Larson1976, Matteucci1989, Pilkington2012}, and radial migration \citep{Schonrich, Minchev2013}. The second episode of metal-poor gas accretion first dilutes the ISM, while subsequent star formation drives the metallicity upward again. At the same time, inside-out growth establishes a radial metallicity structure, and radial migration mixes stars born at different radii. This dynamical mixing broadens the age range over which the dilution signature is observed in the Galactic disk, especially in the outer disk. 
\citet{Matteucci2021} summarized that the Milky Way's thick and thin disks likely formed through distinct gas-infall events, with the thin disk evolving more gradually in an inside-out manner. Radial migration, inward gas flows, and radially varying star-formation efficiencies may have further shaped the disk's abundance gradients and chemical distribution. This picture is broadly consistent with the chemo-dynamical framework adopted in this work.

\begin{figure}
\centering
\includegraphics[width=6.5in]{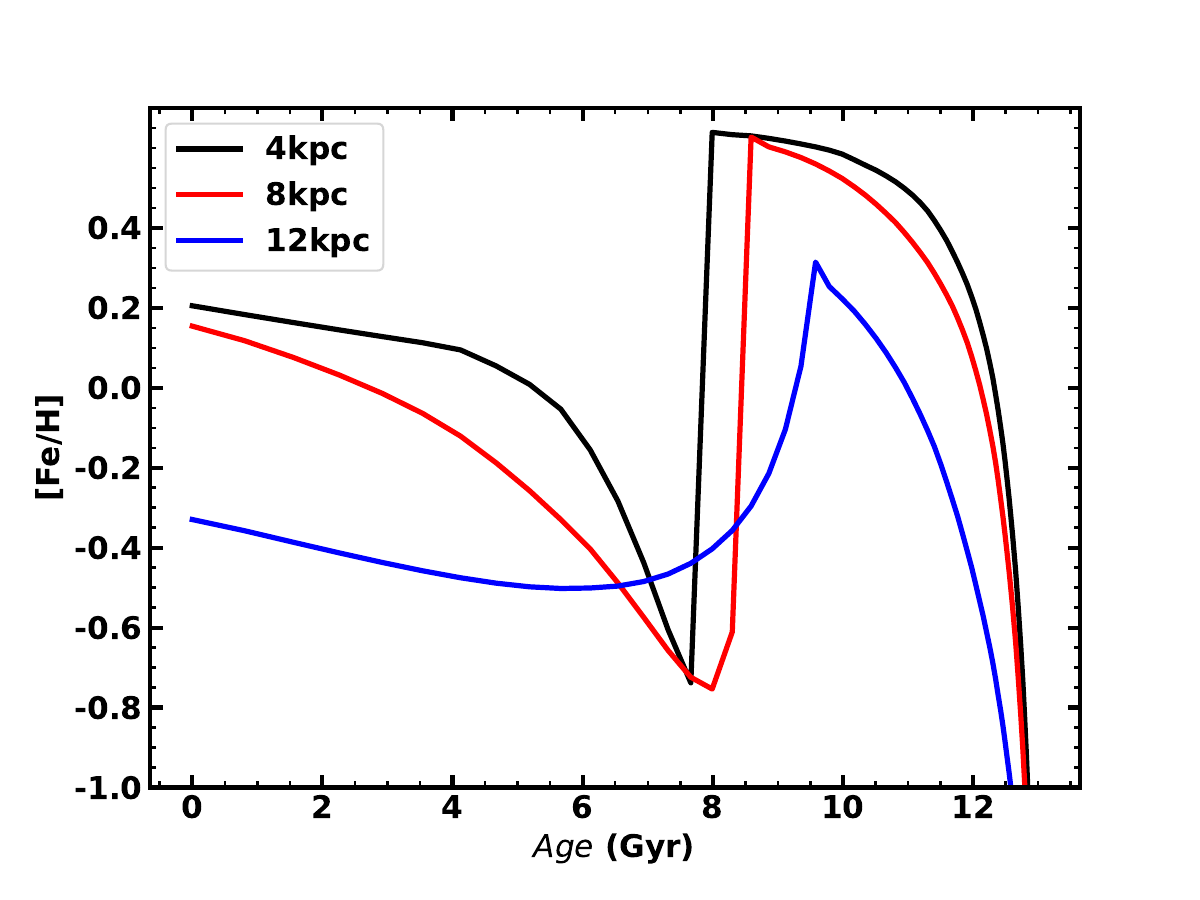}
\caption{ Predicted stellar AMRs from our two-infall models at 4, 8, and 12~kpc. We note that the model does not impose any restriction on the $|Z|$-range, as $|Z|$ is not included as a model parameter.}
\label{age_mass_fe_model}
\end{figure}

To explore this scenario quantitatively, we construct a simple toy model that combines two episodes of gas infall with radial migration. We first build a chemical-evolution model using the JINA-NuGrid chemical-evolution pipeline \citep[JINAPyCEE;][]{Cote2017,Cote2018}. The gas inflow rate is parameterized as an exponential function,
\begin{equation}
    \dot{M}_{\rm gas,in}(t)=N_{\rm norm}\times{exp\left[\frac{-(t-t_{\rm max})}{\tau}\right]},
\end{equation}
where $N_{\rm norm}$ sets the normalized mass of the infalling gas, $\tau$ is the characteristic gas-accretion timescale, and $t_{\rm max}$ denotes the time of maximum infall. We calculate the chemical-evolution histories at Galactocentric radii of 4, 8, and 12\,kpc using the parameters listed in Table~\ref{table1}. The accretion timescales $\tau$ and peak-infall times $t_{\rm max}$ are adopted from \citet{Spitoni2021}. The relative strengths of the two infall episodes, $N_{\rm norm,2}/N_{\rm norm,1}$, together with the corresponding star-formation efficiencies for the two episodes, are also given in Table~\ref{table1}. Figure~\ref{age_mass_fe_model} shows the predicted stellar AMRs from our two-infall models at 4, 8, and 12\,kpc. We note that the models do not impose any restriction on the $|Z|$-range, as $|Z|$ is not included as a model parameter. Compared with \cite{Spitoni2021}, although $t_{\max}$ and $\tau$ are identical between the two models, other model parameters may differ, including parameters not explicitly reported in their paper. Therefore, the detailed predictions are not expected to be identical. Nevertheless, both models exhibit the same overall trend: metallicity increases rapidly following the first gas infall, decreases after the second infall of metal-poor gas, and then gradually increases again as star formation proceeds and the gas becomes progressively enriched.

We then incorporate radial migration following the general approach of \citet{Frankel2018,Frankel2020} and \citet{Spitoni2025}, treating migration as a diffusion process. The probability that a star born at Galactocentric radius $R_{\rm initial}$ is observed at $R_{\rm final}$ after an age $\tau$ is written as
\begin{equation}
    \ln p(R_{\rm final}|R_{\rm initial},\tau)=ln(c)-\frac{(R_{\rm final}-R_{\rm initial})^2}{2\sigma_{\rm RM}^2\tau/10\rm{Gyr}},
\end{equation}
where $\sigma_{\rm RM}$ characterizes the strength of radial migration and $c$ is a normalization constant. To illustrate the effect of efficient migration, we adopt a reference value of $\sigma_{\rm RM}=3.5$\,kpc.

\begin{figure}
\centering
\includegraphics[width=7.5in]{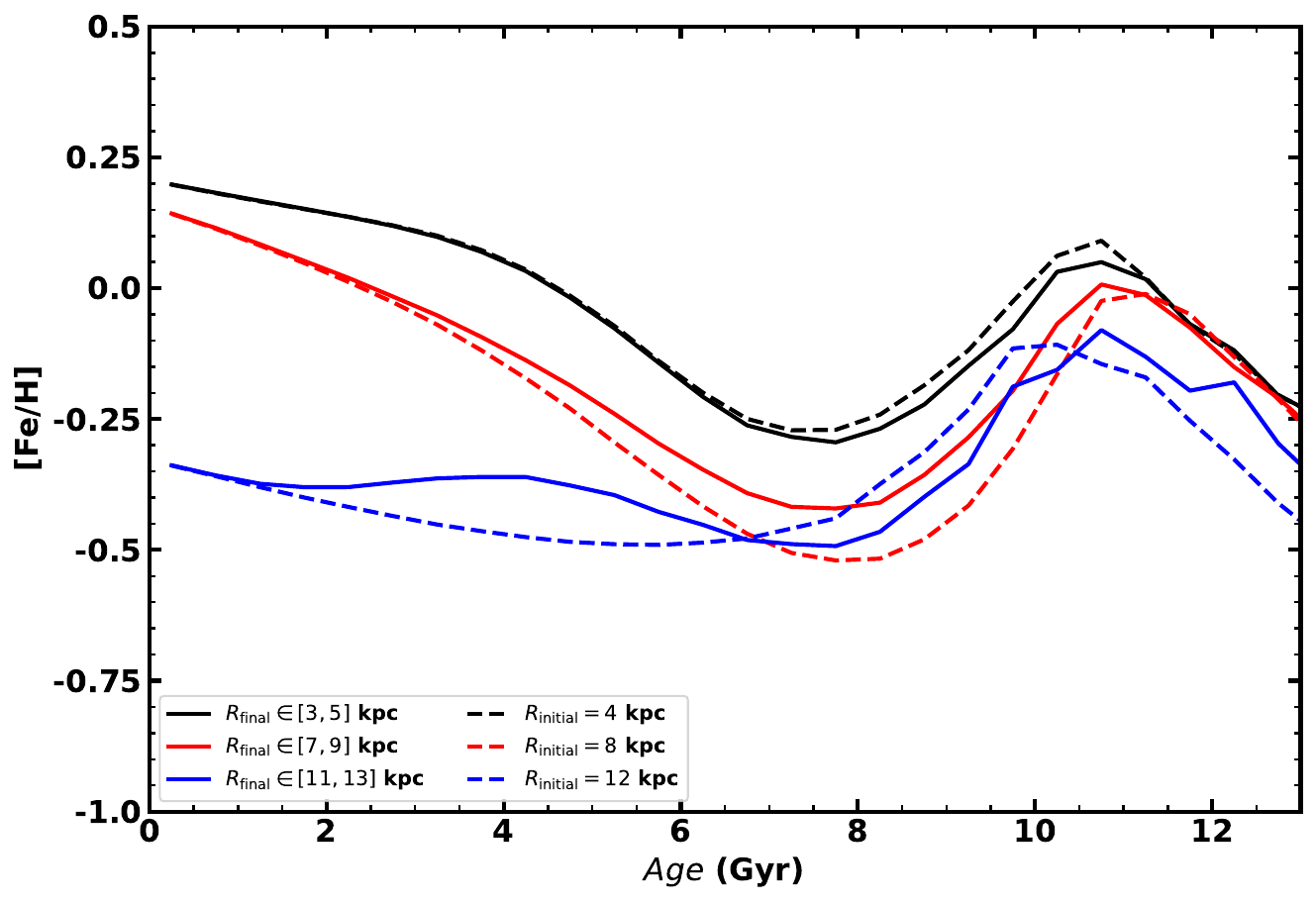}
\caption{Predicted AMRs at 4, 8, and 12\,kpc, with (solid lines) and without (dashed lines) radial migration effects. For comparison with the observations, the model stellar ages are convolved with Gaussian kernels whose dispersions are set by the age uncertainties corresponding to each age bin shown in Figure\,\ref{age-ageerror-relation}.}
\label{amr_predicted}
\end{figure}

\begin{figure}
\centering
\includegraphics[width=6.5in]{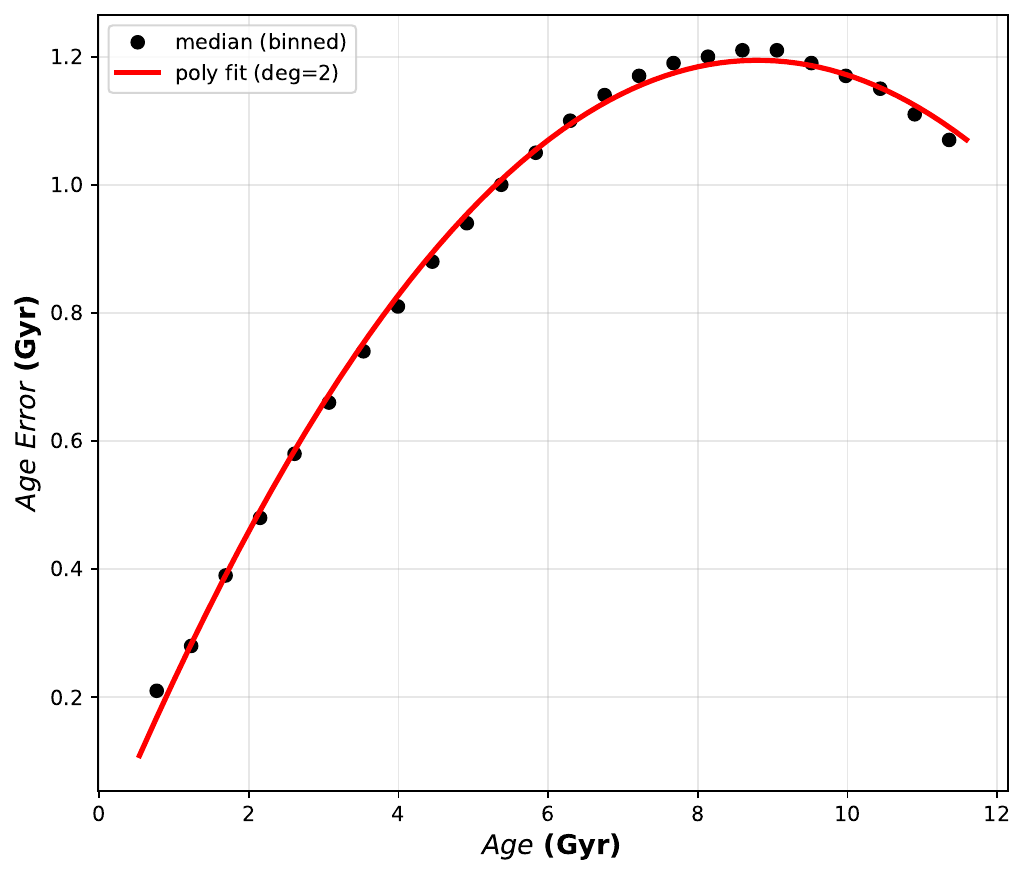}
\caption{Relation between stellar age and age uncertainty based on the observational uncertainties reported by \cite{Anders2023}. The black points represent the median age uncertainties at each stellar age, while the red curve shows the quadratic fit to the relation between stellar age and age uncertainty.}
\label{age-ageerror-relation}
\end{figure}

Figure~\ref{amr_predicted} presents the AMRs predicted by the toy model for two cases: the two-infall chemical evolution  histories alone, and the two-infall histories combined with radial migration. To account for observational age uncertainties, we convolve the model stellar ages with a Gaussian kernel whose dispersion is determined by the age-dependent uncertainties reported by \cite{Anders2023}. For each age, we estimate the median age uncertainty and fit the relation between stellar age and age uncertainty. The fitted relation, shown in Figure\,\ref{age-ageerror-relation}, is then applied to each model stellar age when performing the convolution. As shown in this figure and Figure~\ref{age_mass_fe_model}, both age uncertainties and radial migration smooth the underlying AMR. However, their effects are not identical; age uncertainties mainly blur sharp features, whereas radial migration changes the shape of the relation by introducing stars formed at different Galactocentric radii. Most notably, radial migration produces a metallicity peak at an age of $\sim 3.5\,\mathrm{Gyr}$ in the outer disk. The appearance of this peak, followed by a decline toward the youngest ages, suggests that the AMR in the outer disk  is strongly influenced by stars formed elsewhere in the Galaxy and subsequently redistributed through radial migration, in addition to reflecting the local chemical evolution history.

Our results are broadly consistent with the chemodynamical models of \cite{Minchev2014}, \cite{Johnson2021}, and \cite{Lu2022}, which show that radial migration reshapes the observed AMR by mixing stars formed at different Galactocentric radii. This process broadens the metallicity distribution at a given age and can flatten or enhance features in the AMR, particularly in the outer disc. However, the detailed shape of the AMR, especially the location of its turning points, depends more strongly on the gas accretion history associated with satellite galaxies and on the corresponding star formation histories.

\begin{figure}
\centering
\includegraphics[width=7.5in]{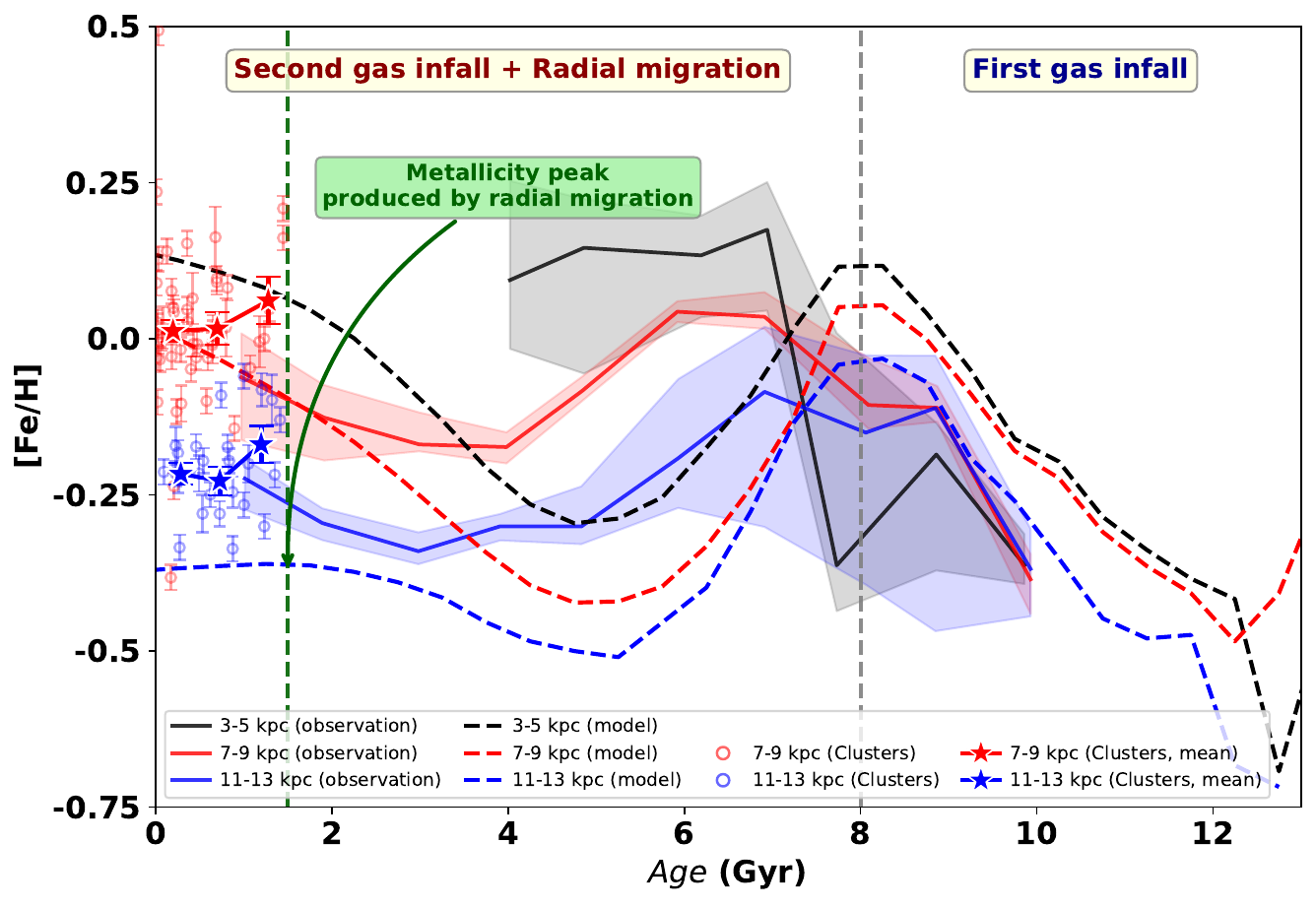}

\caption{The AMRs of APOGEE red clump stars and open clusters in three Galactocentric radial bins,  restricted to $|Z| < 0.5$\,kpc. The AMRs predicted by the toy model, which combines two episodes of gas infall with radial migration, are overplotted from Figure~\ref{amr_predicted}. To account for the systematic offset between the model and observational age scales, all model ages have been shifted by $-2.5\,\mathrm{Gyr}$. The shifted model predictions and APOGEE red clump observations are shown in black, red, and blue for the radial bins $3<R<5\,\mathrm{kpc}$, $7<R<9\,\mathrm{kpc}$, and $11<R<13\,\mathrm{kpc}$, respectively. The model predictions are represented by dashed lines, while the observations are shown by solid lines with shaded regions.  Open cluster measurements are shown as points, while the mean AMR for the open clusters is indicated by star symbols for bins containing more than five stars. The dominant physical processes over different age ranges are also indicated. The gray dashed line marks the approximate onset time of the second gas infall episode, while the green dashed line marks the metallicity peak produced by radial migration in the outer disk. }
\label{radial_migration}
\end{figure}

In Figure~\ref{radial_migration}, we compare the AMRs predicted by the toy model with the observed relations. By comparing the results shown in Figures~\ref{age_mh} and~\ref{amr_predicted}, we find that the model ages are systematically older than the observed ages by approximately $2.5~\mathrm{Gyr}$. Applying a uniform shift of $-2.5~\mathrm{Gyr}$ brings the model into good agreement with the observed trend. This empirical correction likely reflects systematic differences between the model and observational age scales; therefore, the shifted model is used for comparison rather than as a precise prediction of the absolute ages. We also show the AMRs of open clusters, using their ages and Galactocentric distances from \cite{Donor2018}, and \cite{Cantat-Gaudin2020} and metallicities from \cite{Donor2018}, and \cite{Otto2026}. The sample includes 19 clusters from \cite{Donor2018} and 158 clusters from \cite{Cantat-Gaudin2020}. Since the former catalogue does not provide $Z$ values, all 19 clusters are included, whereas the latter sample is restricted to clusters with $|Z| < 0.5\,\mathrm{kpc}$. Both the observed and predicted relations are non-monotonic. For stars older than $6$--$8\,\mathrm{Gyr}$, the AMR is mainly shaped by the first episode of gas infall and the associated star formation.  At younger ages, the delayed second infall introduces relatively metal-poor gas into the ISM, causing the metallicity of newly formed stars to decline, while subsequent star formation and chemical enrichment drive a partial recovery toward the youngest ages. This sequence of events naturally explains the observed non-monotonic trend: a metallicity decline from $\tau \sim$ 6--8\,Gyr to younger populations,  reaches a minimum at intermediate ages, and then rises again.   In the outer disk, however, the model including radial migration predicts an additional metallicity peak at an age of $\sim 2.0\,\mathrm{Gyr}$.  This peak indicates that the outer-disk AMR undergoes a renewed reversal: the metallicity first increases toward younger ages, reaches a maximum, and then declines again for the youngest populations. Such a localized peak cannot be naturally produced by the standard two-infall scenario alone, which can explain the initial dilution and subsequent chemical recovery but not the additional turnover at an intermediate age. Instead, the peak likely reflects the redistribution of metal-rich stars born in the inner disk through radial migration, while the youngest stars, having experienced less migration, more closely preserve the lower metallicity of the local outer disk.  The peak is not prominent in the AMR of APOGEE red clump stars, possibly because stars younger than 2\,Gyr have larger relative age uncertainties and less reliable age estimates. The AMR of the open clusters, in contrast, shows a clearer peak at an age of $\sim$\,1.5-2.0\,Gyr. Despite the difference in the exact age of the peak, our results indicate that radial migration plays a crucial role in shaping the outer-disk AMR. 
%{In the outer disk, particularly at $11<R<15$\,kpc, radial migration significantly modifies the age--metallicity relation and extends the regime in which metallicity decreases toward younger ages. In contrast, at $R\lesssim7$--9\,kpc, the influence of radial migration on the predicted age--metallicity relation is comparatively weak. This radial dependence is qualitatively consistent with Figure~\ref{age_mh}, where the declining branch of the observed age--metallicity relation is most prominent in the outer disk.}

Although this toy model is intentionally simplified and is not intended to provide a unique reconstruction of the Milky Way's chemical evolution history, it reproduces the main qualitative features of the observations. The second episode of metal-poor gas accretion produces the metallicity dilution, subsequent chemical enrichment drives the recovery toward the youngest populations. The observed non-monotonic AMR is unlikely to be the imprint of any single process; instead, it reflects the interplay between episodic gas accretion, continued chemical enrichment, inside-out disk growth, and radial migration. The distinctive metallicity peak ($\sim$\,1.5-2.0\,Gyr) and subsequent turnover in the outer disk AMR provide particularly strong evidence that radial migrations have played an essential role in shaping the chemical structure of the Milky Way disk. 

\begin{table}[]
    \centering
    \begin{tabular}{|c|c|c|c|c|c|c|c|}
         \hline &$N_{\mathrm{norm2}}/N_{\mathrm{norm1}}$ & $t_{\mathrm{max},1}$ [Gyr] & $\tau_{1}$ [Gyr] & $\mathrm{SFE}_{1}$ & $t_{\mathrm{max},2}$ [Gyr] & $\tau_{2}$ [Gyr] & $\mathrm{SFE}_{2}$ \\
         \hline
            4\,kpc &3.0 & 0&0.115 &3.0 &4.6 & 3.75& 1.5 \\
         \hline
            8\,kpc & 5.0& 0& 0.103& 2.0&4.0 &4.11 &0.5 \\
         \hline
            12\,kpc &5.0 &0 & 0.449&2.0 & 3.05&10.98 & 0.1 \\ 
        \hline  
    \end{tabular}
    \caption{We present the parameters for our models at 4 kpc, 8 kpc, and 12 kpc (refer to the text for model details). The second to fifth lines detail the parameters of the first gas infall, while the sixth to eighth lines cover the parameters of the second gas infall.}
    \label{table1}
\end{table}

%We examine the impact of radial migrations on the age-metallicity relation, employing the same methodology as in \cite{Spitoni2025} and \cite{Frankel2018}, where migration is viewed as a diffusion process treated parametrically \citep[as noted in][]{Frankel2020}. Specifically, the probability of a star being found at a Galactocentric radius $R_{\rm final}$, given its birth at $R_{\rm initial}$ and a specific age ($\tau$), can be expressed as:
%\begin{equation}
%    ln p(R_{final}|R_{initial},\tau)=ln(c)-\frac{(R_{final}-R_{initial})^2}{2\sigma_{RM}\tau/10\rm{Gyr}}
%\end{equation}
%Here, $\sigma_{RM}$ (in kpc) denotes the strength of radial migration, and $c$ is a normalization constant that prevents stellar migration to negative radii. To investigate the effects of extreme migration, we adopt a reference value of $\sigma_{RM} =$6.0\,kpc. 

% Figure \ref{radial_migration} displays the age-metallicity relationships \textbf{predicted by a toy model combining two gas-infall episodes and radial migration processes. From the figure, we can find that metallicity decreases with age for $\tau < 4 \, \text{Gyr}$ at $11 < R < 15$\,kpc, indicating that radial migration contributes to the decline in metallicity with decreasing age shown in Figure \ref{age_mh}.  However, for stars at $R < $7--9\, \text{kpc}, the effect of radial migration on the predicted age-metallicity relationship is negligible.} Figure \ref{age_mh} illustrates that the decrease in metallicity with decreasing age is more pronounced in the outer disk, aligning with the predictions of the toy model.  

%In conclusion, the Galactic disk experienced two main episodes of gas infall, leading to the formation of the halo-thick disk and the thin disk. The second infalling gas  is chemically pristine, thus old stars ($\tau \sim $\,8--9\,Gyr) formed primarily from the unenriched intergalactic medium  are predicted to be more metal-poor than those formed earlier from the enriched thick disk gas. The radial migration process produce the metallicity with decreasing age for young stars ($\tau <\,$6--8\,Gyr). 

\section{Summary}

In order to quantify the importance of radial migration and gas infall processes in the Galactic disk formation and evolution, we use APOGEE red clump stars and open clusters with precise age estimates to map the AMRs across the Galactic disk and at different Galactocentric radius $R$ and vertical height $Z$. 

For stars older than approximately 6--8 Gyr, the AMR is nearly independent of $R$ and $Z$ within the measurement uncertainties. This weak spatial dependence suggests that the early disk experienced a rapid and relatively well mixed chemical enrichment, with different regions evolving in an approximately self-similar manner.
In contrast, younger stellar populations of $\tau < $6--8\,Gyr exhibit a clear negative radial metallicity gradient, with the inner disk being systematically more metal-rich than the outer disk. This indicate an ''inside-out" disk formation scenario, in which star formation and chemical enrichment proceeded more rapidly in the inner Galaxy than at larger radii.

%Most importantly, the AMR for stars younger than approximately 6--8 Gyr is distinctly non-monotonic. At fixed Galactic location, metallicity first decreases toward younger ages, reaches a minimum at intermediate ages, and then increases again among the youngest stars. In addition, the outer disc ($R > 11\,\mathrm{kpc}$) exhibits a metallicity peak at an age of $\sim 2\,\mathrm{Gyr}$. These features cannot be explained by continuous local chemical enrichment alone.  Instead, it points to competition between metal-poor gas accretion, which temporarily dilutes the interstellar medium, and subsequent star formation and enrichment, which raise the metallicity again. Radial migration further broadens and reshapes the observed relation, particularly the metallicity peak at $\sim 2\,\mathrm{Gyr}$ in the outer disc ($R>11\,\mathrm{kpc}$), by redistributing stars formed at different Galactocentric radii.

Most importantly, the AMR for stars younger than approximately $6$--$8$\,Gyr is clearly non-monotonic. At a fixed Galactocentric radius, the metallicity first decreases toward younger ages, reaches a minimum at approximately $3$--4Gyr, and then increases again among the youngest stars. In the disc of $R>$7\,kpc, the metallicity appears to decline again at around $1.5$--$2.0$\,Gyr. This behavior cannot be explained by continuous local chemical enrichment alone. Instead, it likely reflects the competition between dilution by the accretion of metal-poor gas during the second gas infall episode and subsequent enrichment driven by star formation. The initial decrease in metallicity from approximately 6--8\,Gyr ago may have resulted from the accretion of metal-poor gas, whereas the subsequent increase after approximately 3--4\,Gyr ago reflects the increasing importance of star formation driven enrichment. The later decline over the past approximately 1.5-2.0\,Gyr, particularly that seen in the open cluster sample, may be associated with radial migration, which can further broaden and reshape the AMR by redistributing stars born at different Galactocentric radii.

To interpret this non-monotonic AMR for young stellar populations ($\tau<$8\,Gyr), we construct a toy model with JINAPyCEE chemical evolution coupled to radial migration. We find that the observed behavior can be reproduced by the combined effects of (i) two major gas infall episodes and (ii) radial migration. In this picture, the second infall gas is relatively metal-poor and dilutes the ISM, producing a minimum in metallicity at $\tau\sim$8--9 \, Gyr relative to stars formed earlier from a more enriched ISM; radial migration broadens and enhances the declining branch and produces a metallicity peak at an age of $\sim$\,1.5-2.0\,Gyr in the outer disc.                                                                                                                                                                                   

\bibliography{age_metallicity_relation.bib}{}
\bibliographystyle{aasjournal}


\begin{acknowledgments}
 This work was funded by the National Key R\&D Program of China
(No. 2019YFA0405500) and the National Natural Science Foundation of China (NSFC Grant Nos. 11988101, 12422303, 11833006, 11973001, 12173007, 12422303, and 12273053).
Guoshoujing Telescope (the Large Sky Area Multi-Object Fiber Spectroscopic
Telescope LAMOST) is a National Major Scientific Project built by the Chinese Academy of Sciences.
Funding for the project has been provided by the National Development and Reform Commission.
LAMOST is operated and managed by the National Astronomical Observatories, Chinese Academy of
Sciences. The LAMOST FELLOWSHIP is supported by Special Funding for Advanced Users, budgeted and administrated by Center for Astronomical Mega-Science, Chinese Academy of Sciences (CAMS). Supported by High-performance Computing Platform of Peking University.  T.C.B. acknowledges partial support from Physics Frontier Center/JINA Center for the Evolution of the Elements (JINA-CEE), and OISE-1927130: The International Research Network for Nuclear Astrophysics (IReNA), awarded by the US National Science Foundation (NSF).
\end{acknowledgments}

\end{document}